# Evaluation of High-Speed Universal Shift Register with 4-bit ALU

Md Shahriar Kabir[1], Khalid Mahmud Niloy[1], Sumaiya Afrose[1]
[1]*Dept. of Electrical & Electronics Engineering*
[1]*American International University-Bangladesh*
Email- tulip6585@gmail.com, kmniloy1999k@gmail.com, sumaiyaafrose085@gmail.com

***Abstract*—This paper contains information about the universal shift register. In the early stages of this paper, this paper introduces different types of flip flops and calculates the delay. After that, different types of flip flops are used to make a universal shift register, and the high-speed universal shift register is measured using a timing diagram. In addition, a complete memory system was designed at the end of this paper. A universal shift register with 4-bit Alu was added to complete the memory system. As a result, this method has created an accurate memory storage device with high-speed characteristics.**

Keywords—***Universal Shift Register, 4-bit ALU, Full adder, 4*1 Multiplexer, TGMS D flip-flop.***

## I. Introduction

.Flip flops can store or hold a single bit of binary data (0 or 1). However, several flip-flops are required to store several bits of data. The register is a device that stores such information. A flip-flop array is a set of flip-flops to hold multiple data bits. The data contained in these registers can be transferred or shifted via shift registers. By using clock pulses, the bits contained in such registers can change within and in/out of the registers. A high-speed shift register can shift data at a low latency, which is beneficial and desirable. It can attain extremely little latency in data processing by using an emerging notion of high speed in VLSI architecture. As a result, high-speed design can increase the performance of all preceding design stages. A system can have two types of timing errors: hold time violations and setup time violations. A hold time violation occurs when an input signal changes too soon after the clock's active transition. In a setup time infraction, the signal arrives too late and misses the time because it should advance. The arrival time of the signal can vary related to variations in input data, circuit activities, temperature and voltage variations, and so on.

High-speed shift registers require high-speed flip-flops. Therefore, this project has tested various flip-flops, explored high-speed flip-flops, and compared them. The lower the universal shift register's data shift delay, the higher the speed. ALU has also been designed and added to the universal shift register to complete the system and observe different ALU microoperations during the data shifting.

## II. Literature Review

Rajaram et al. have designed a system of high-speed shift registers using a single clock pulse method. This study proposes a new shift register design based on a single clock pulse with Hold Mode (HM-FF) and without Hold Mode (WHM-FF) Flip Flop. This approach is designed for the Xilinx Virtex 6 family and delivers a reasonable speed boost. Compared to the present method, it gives a 41.9 percent reduction in latency. Future work is related to power optimization and area overhead of the implementation using this design [1].

Praveen et al. have invented 2GHz High-Performance Double Edge Triggered D-Flip Flop Based Shift Registers In 32NM CMOS Technology. Data is stored on a clock signal's rising and falling edges in Double Edge Triggered Flip Flops. Although system specifications control the clock frequency, DET flip-flops can reduce the clock frequency to half its original value while maintaining the same data throughput. As a result, DET flip-flops consume less power, making them ideal for low-power applications. The performance of the DETFF-based shift registers is evaluated by analyzing the average power, latency, and PDP at various clock frequencies [2].

## III. Basic Idea of Project

Fig. 1 contains a universal shift register, where S0 and S1 are the selected pins used to choose the mode of operation, shift proper operation, or parallel mode. Pin-D of the first 4×1 Mux is fed to the output pin of the first flip-flop. Pin-C of the first 4X1 MUX is connected to serial input for shift right. In this mode, the register shifts the data towards the right. Similarly, pin-B of 4X1 MUX is connected to the serial input for shift-left. The universal shift register shifts the data towards the left in this mode. M1 is the parallel input data given to pin-3 of the first 4×1 MUX to provide parallel mode operation and store the data in the register. Similarly, the remaining parallel input data bits are given to the pin-3 of related 4X1MUX to provide parallel loading. F1, F2, F3, and F4 are the parallel outputs of Flip-flops, which are associated with the 4×1 MUX [3].

In Fig. 2, a 4-bit ALU is connected with the universal shift register to look like a memory system. As a result, it can perform eight micro-operations and store data with the help of the universal shift register, which is the central theme of this project.

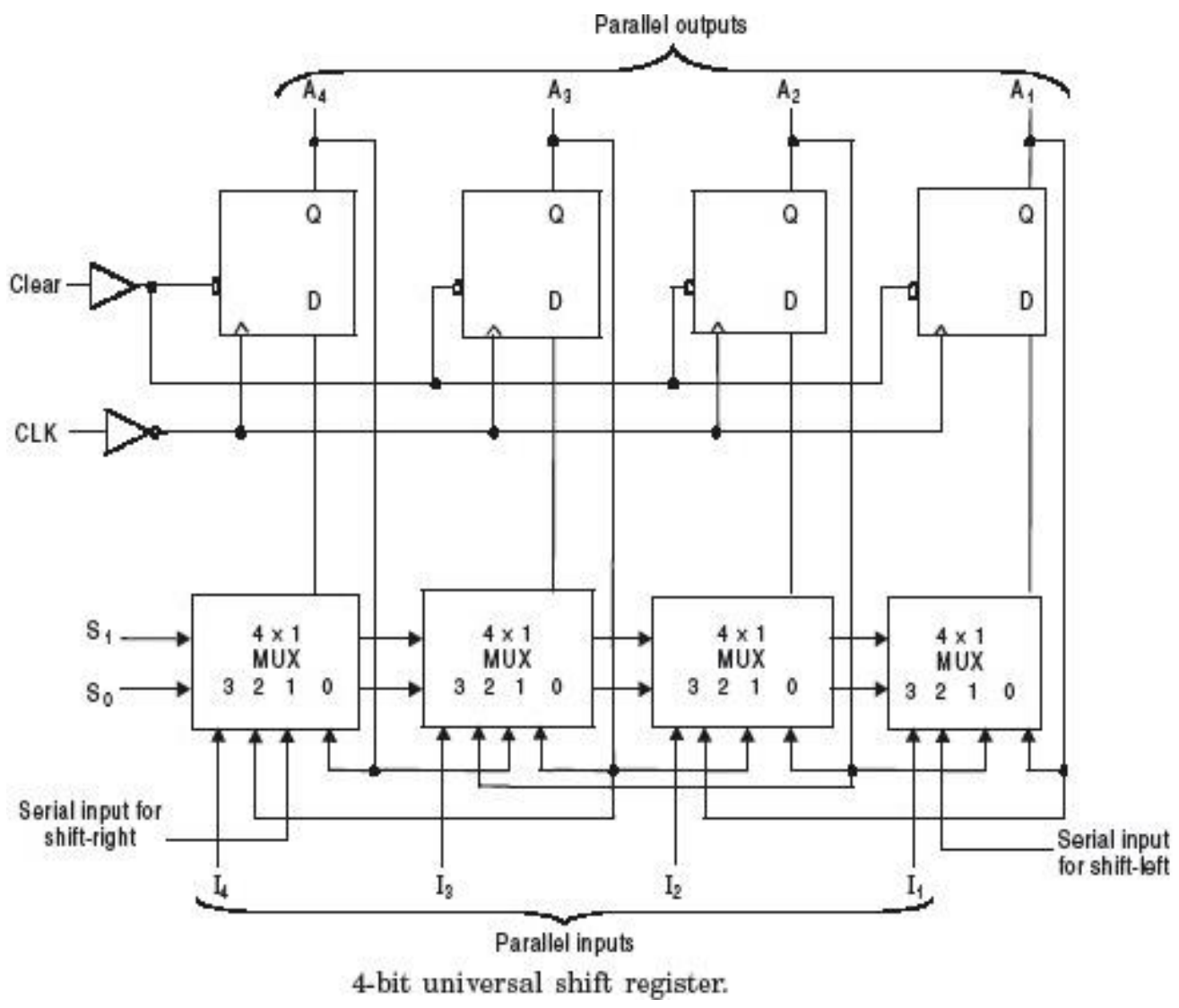

Fig. 1. Universal Shift Register

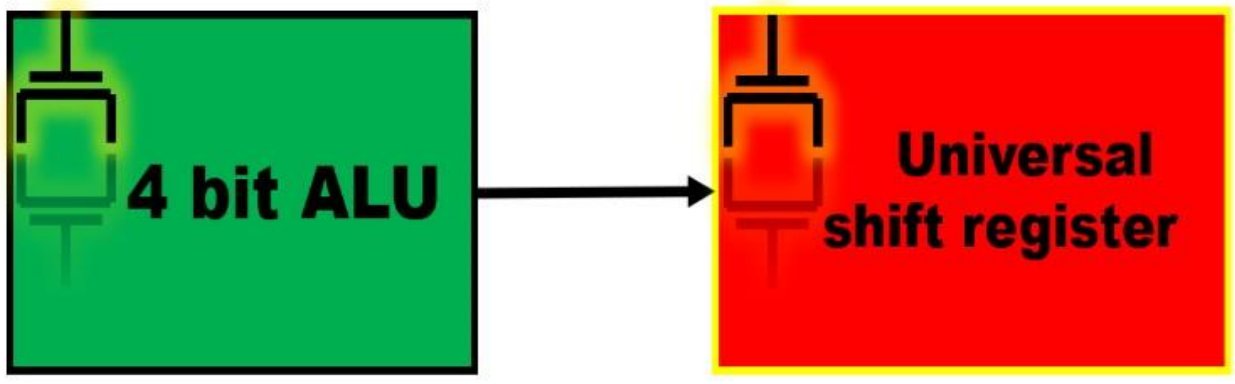

Fig. 2. Schematic of 4-bit ALU with Universal Shift Register

## IV. Software Implementation for Water Level Scale Sensor

A Universal shift register is bidirectional. It is a register with both right and left shift and parallel load capabilities. As a result, all shift register operations are possible in a single circuit. There are flip-flops for designing the universal shift register, and a 4*1 multiplexer is required. A multiplexer is a combinational logic circuit that generates the output from only one input at a time. The select input lines decide which signals govern which input will be reproduced at the output end. Four-to-one (4:1) MUX is used when there are four inputs, and only one is selected to link with the output. Multiplexer controls the operation of the universal shift register according to the below truth table:

Table 1: Truth table for 4*1 multiplexer

| S1 | S0 | Operations |
|---|---|---|
| 0 | 0 | No Change |
| 0 | 1 | Shift Right |
| 1 | 0 | Shift Left |
| 1 | 1 | Parallel Load |

Here, S1 and S0 are the control pins of the multiplexer. When both become 0,0, path 0 (D) is connected to the output. Also, for 0,1, a path 1 (C) is connected to the production, for 1,0, a path 2(B) is connected to the output, and for 1,1, a path 3(A) is connected to the production. According to the circuit in no change operation, the data rotates in a particular cycle and will continue to show the same initial value. In the proper operation of the shift, the data will continue to shift from left to right one by one for the per-clock pulse. In shift left operation, the data will change from right to left one by one for the per-clock pulse. In parallel load operation, all bit data will be loaded simultaneously and provide output by a single clock pulse. Now, it is time to see how flip-flop works. Flip-flop works according to the below truth table:

Table 2: Truth table for Flip-Flop

| Clk | D | Qn |
|---|---|---|
| 0 | x | Qn |
| 1 | 0 | 0 |
| 1 | 1 | 1 |

The truth table shows that when data is provided according to the data, the output is stored. Different types of flip-flops will be discussed to evaluate high speed.

### *A. D Flip-Flop based Universal Shift Register*

The D flip-flop is an edge-triggered device that transfers input data to Q when the clock's edge rises or falls. The D flip-flop is a timed or clocked flip-flop with a single digital input denoted by the 'D.' When a D flip-flop is timed or clocked, its output corresponds to the state of 'D.' D and Clk are the only inputs on the D Flip Flop. The D inputs are accurately routed to the S input, and its complement is routed to the R input. The D flip flop obtains the destination from its ability to manage data in its internal store. The D flip-flop-based shift register is designed according to the universal shift register circuit. The parallel loading and output are shown in Fig. 4.

### *B. Dynamic TGMS-based Universal Shift Register*

TGMS stands for transmission-gate-based master-slave. This work evaluates the dynamic TGMS or D-TGMS flip-flop to attain high speed. However, it is susceptible to clock overlap. Transmission gates and MOS transistors are used to build the DTGMS flip-flop. It has a lower delay than other flip-flop topologies. Furthermore, the energy per transition and clock energy of D-TGMS are also lower than those of NAND, mC2MOS, and C2MOS. This flip-flop fails if the clocks overlap for an extended amount of time. The detailed operation of D-TGMS is shown in Figures 5 & 6 [4]. According to the universal shift register circuit, the TGMS-based shift register has been designed. The parallel loading and output are shown in Fig. 6.

### *C. MTSPC double-edge triggered flip flop-based Shift Register*

MTSPC stands for modified true single-phase clocked. An accurate Single-Phase Clock (TSPC) is a high-speed universal dynamic flip-flop. To halt the toggling of the intermediate nodes, the MTSPC D flip-flop requires one more PMOS than the TSPC. It is commonly used in digital design. According to the proposed MTSPC DFF architecture, pre-charging node B should be halted whenever the path to the ground is ON to avoid toggling. A simple solution that works, in this case, is to add a PMOS transistor that stops the pre-charging phase from occurring without compromising the flip-overall flop's functionality. The proposed MTSPC DFF consumes less power and has a greater maximum oscillation frequency for high performance [5]. The MTSPC flip-flop-based shift register is designed according to the universal shift register circuit.

The parallel loading and output are shown in Figure 8. The MTSPC flip-flop-based shift register is designed according to the universal shift register circuit. The parallel loading and production are shown in Fig. 8.

### *D. TGMS D Flip-Flop-based Universal Shift Register*

TGMS stands for transmission-gate-based master-slave, but it is not dynamic. Fig. 9 depicts a Transmission Gate Based Master-Slave D Flip Flop. The TGs T1 and T2 serve as latches in the master and slave sections, respectively, while an inverter generates inverted and non-inverted clock signal CLK and CLKB locally. As indicated in Fig 10, feedback is delivered from the output node to a specific internal node in the master stage to make the flip-flop static. This feedback is used with the knowledge that the forward path has exactly two inversions. Compared to traditional static designs, the feedback approach used in the design is entirely different. This design decreases the number of transistors and TGs in the critical path, resulting in maximum performance and symmetrical delays [6].

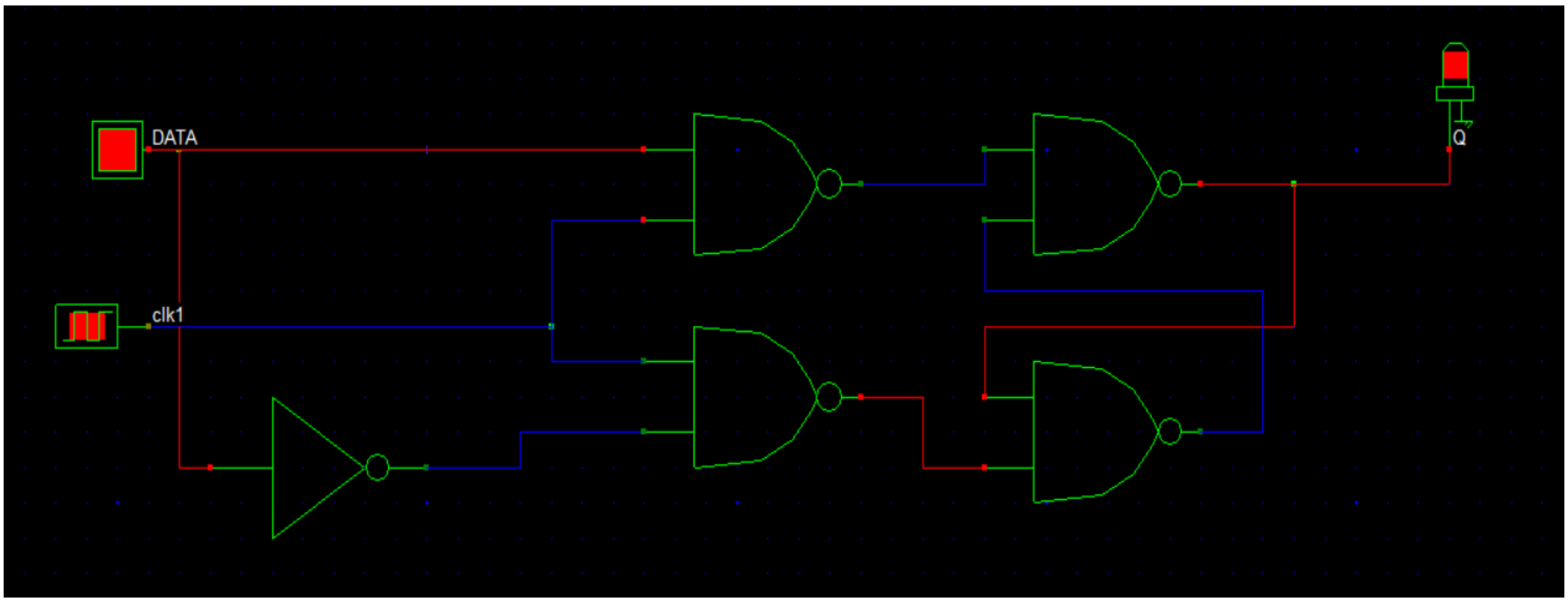

Fig. 3. D Flip Flop

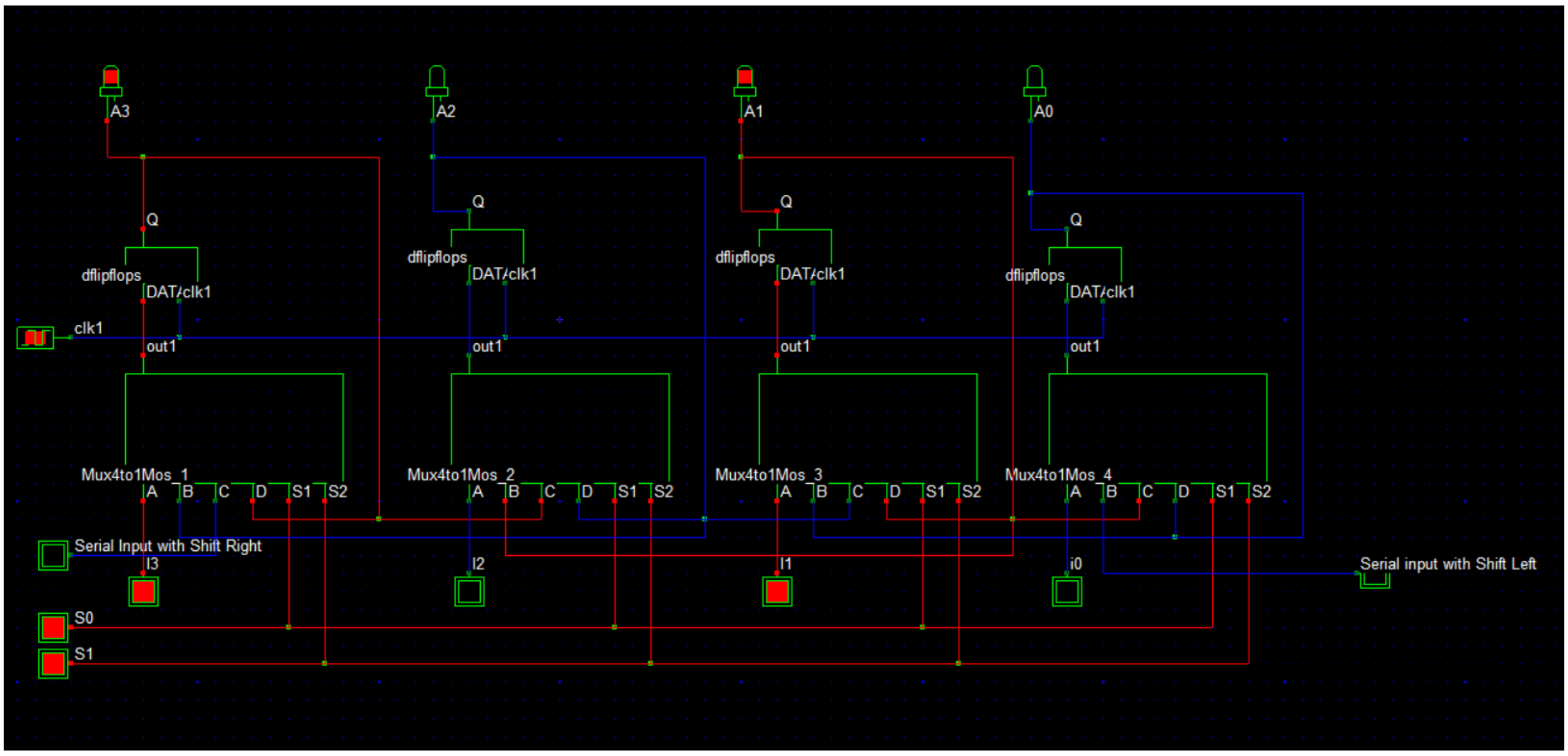

Fig. 4. D flip flop-based Universal Shift Register

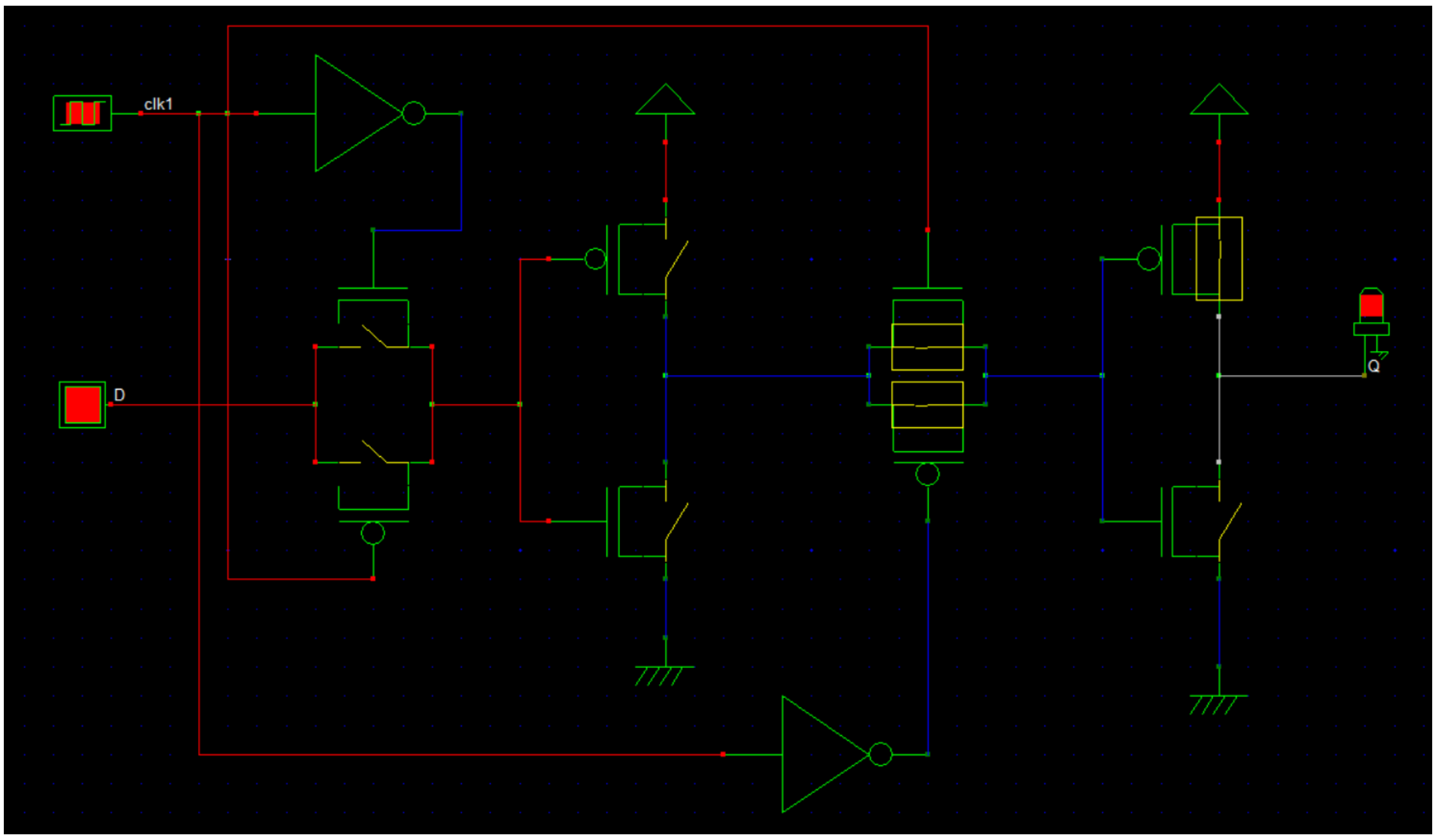

Fig. 5. Dynamic TGMS Flip Flop

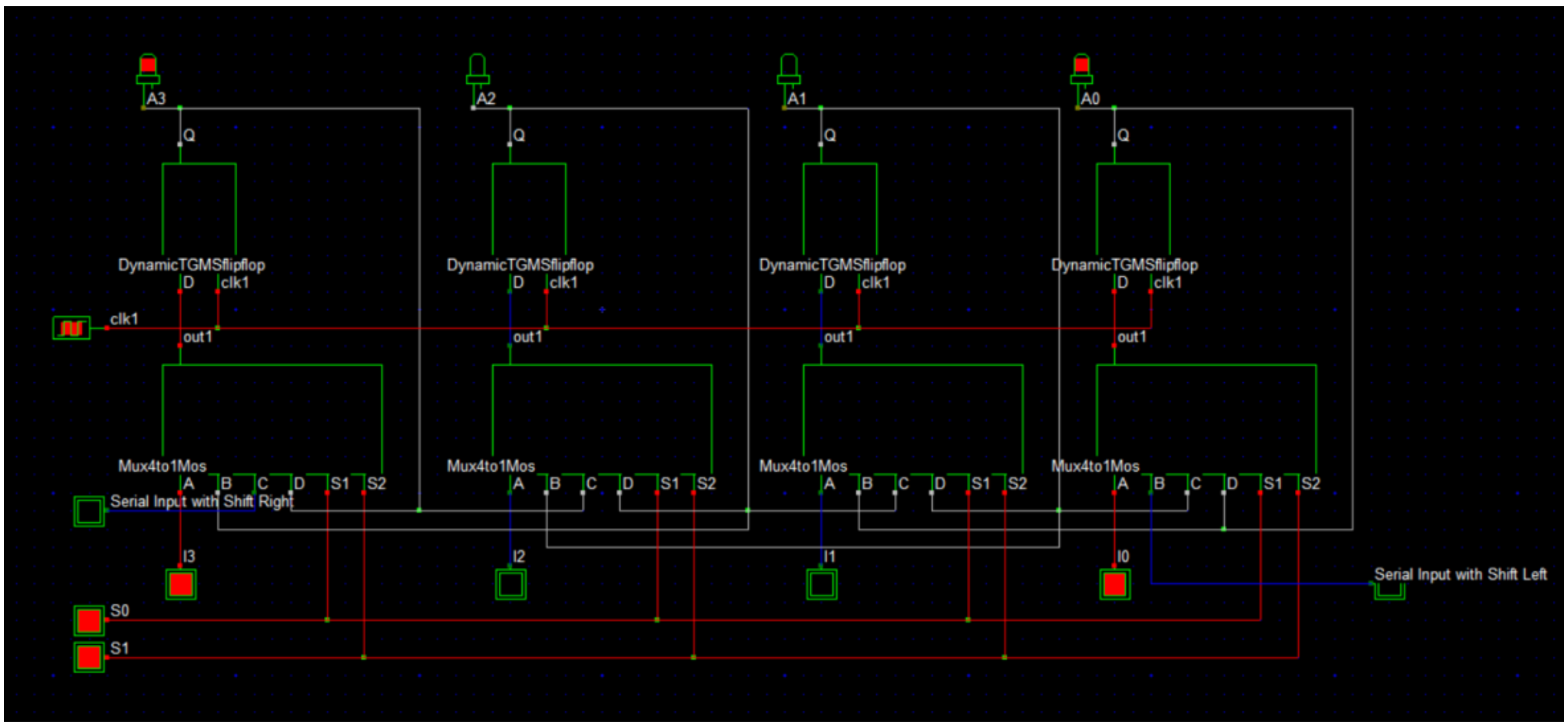

Fig. 6. Dynamic TGMS-based universal shift register

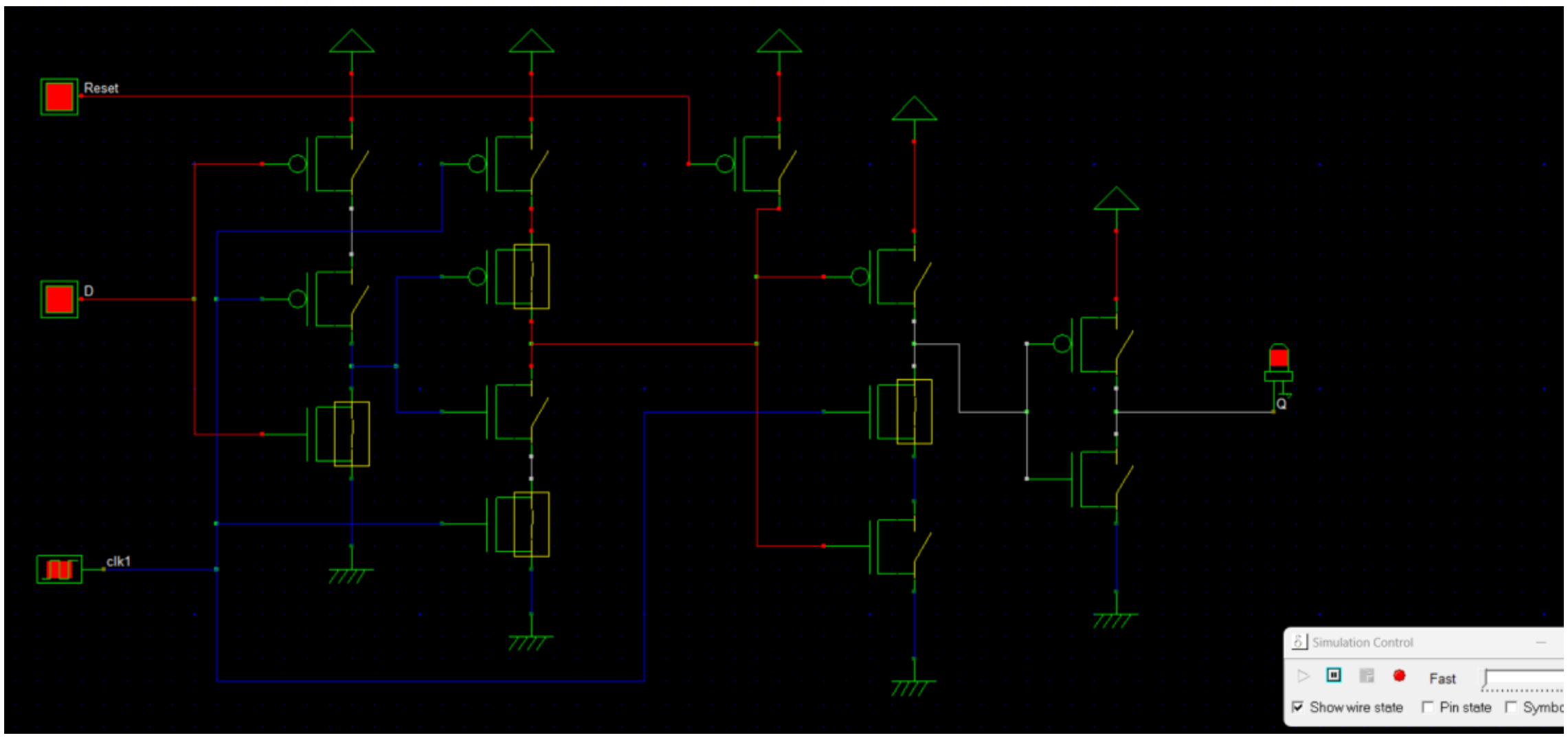

Fig. 7. MTSPC double-edge triggered flip-flop.

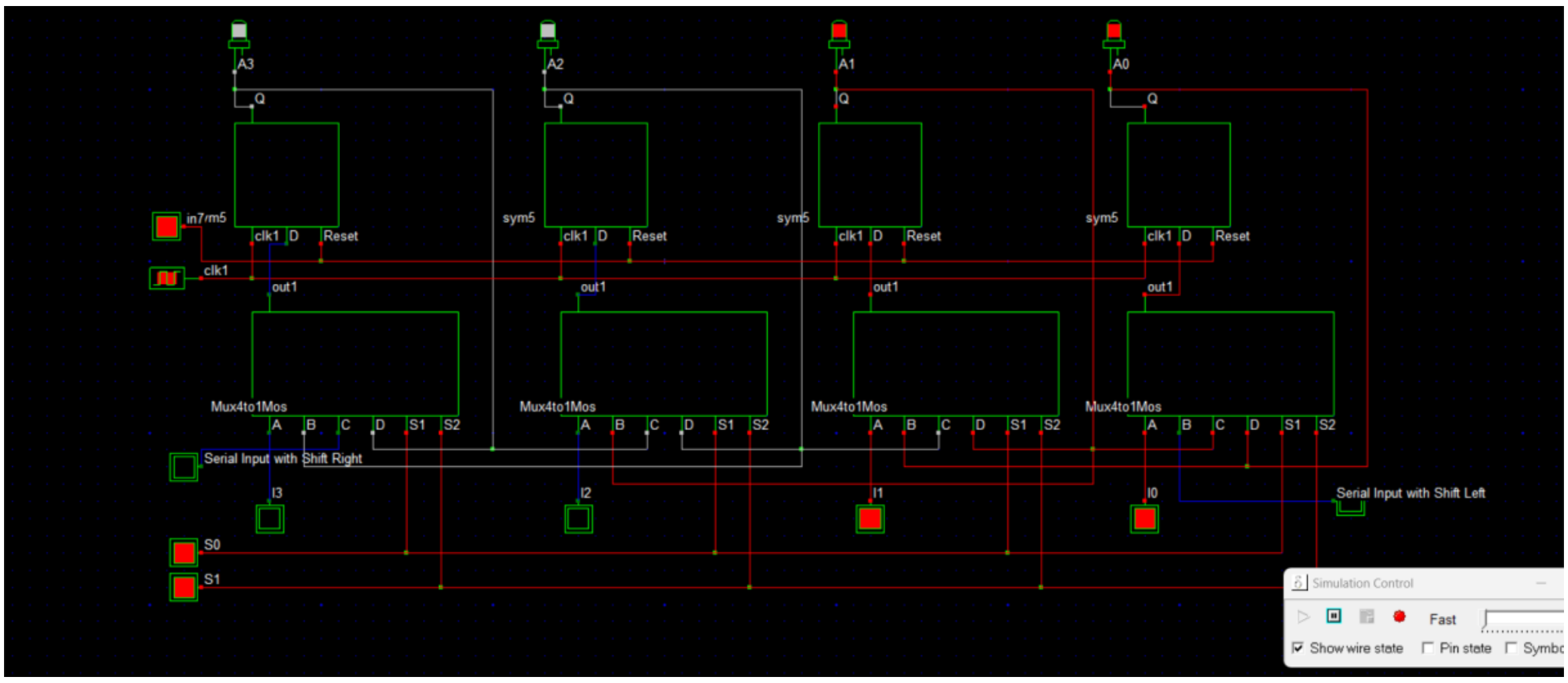

Fig. 8. MTSPC-based Universal Shift Register.

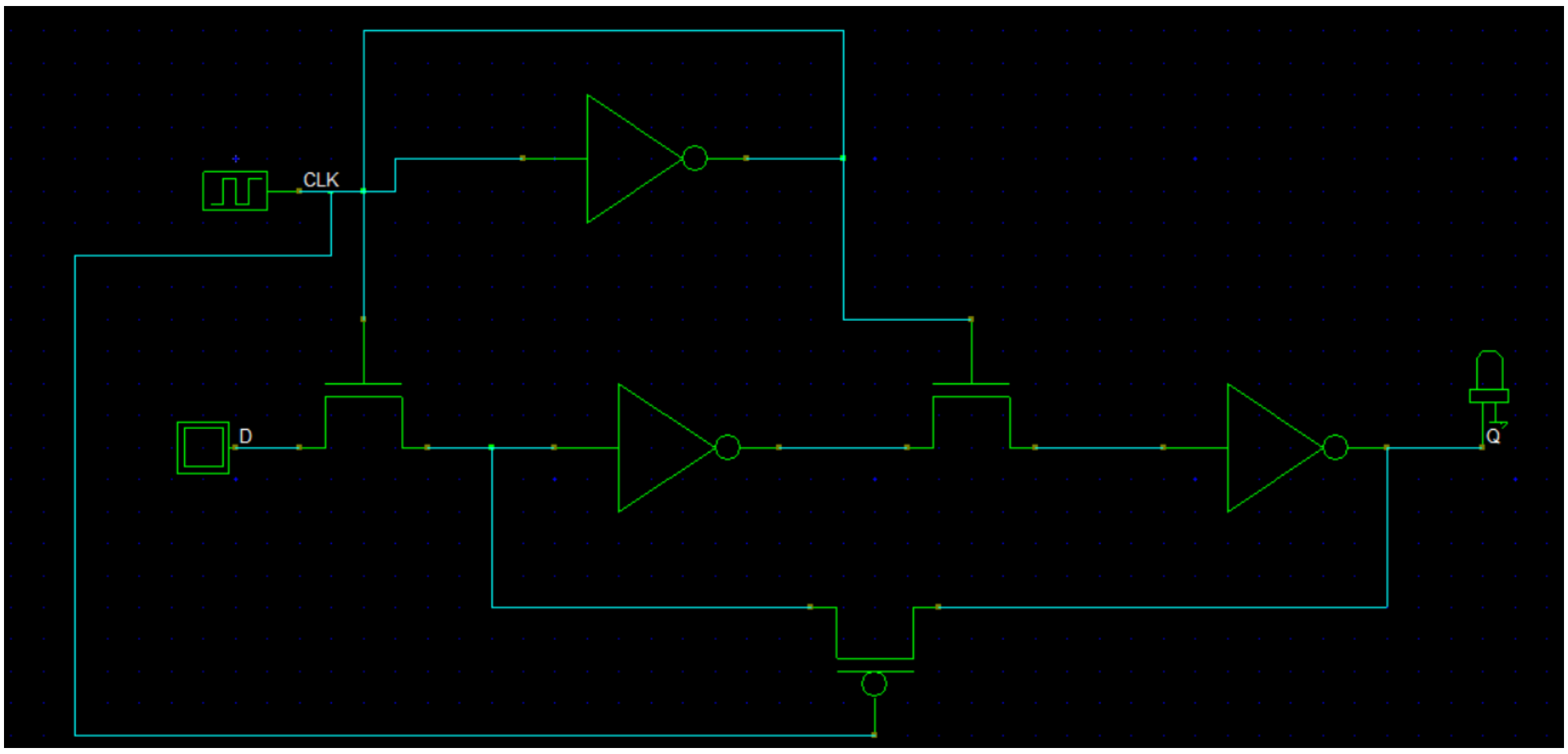

Fig. 9. TGMS D Flipflop

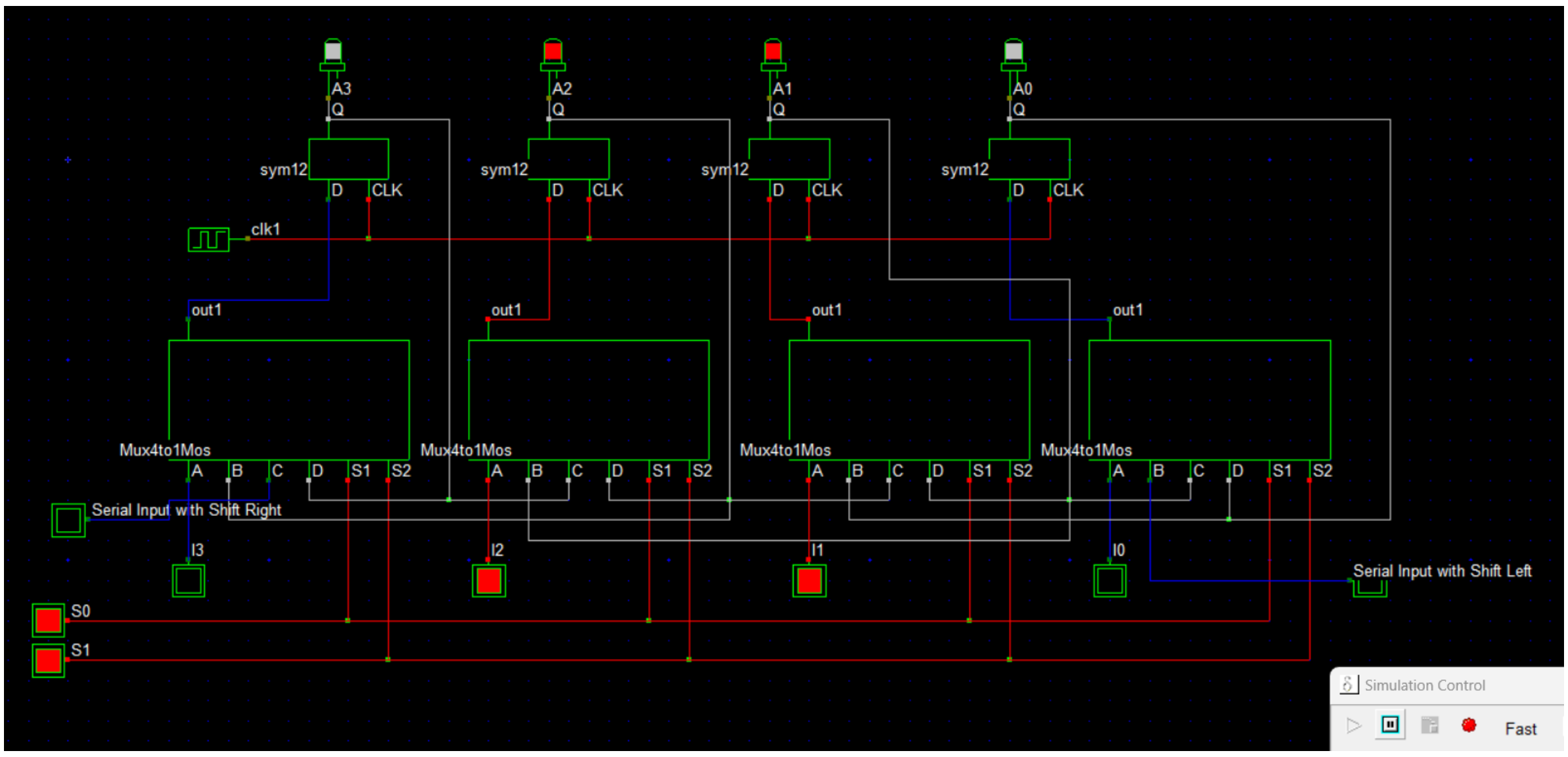

Fig. 10. TGMS D flipflop-based Universal Shift register.

## V. Software Implementation for Universal Shift Register with 4-bit ALU

An arithmetic-logic unit is a portion of a central processing unit that performs arithmetic operations and logic operations on the operands of computer instruction words. The ALU has direct input and output access to the processor controller, main memory (RAM in a personal computer), and input/output devices. [4] So, in this portion of the project, an attempt has been made to design a complete system. This design will enable some microoperations in the ALU and then see how the data shifted. In 4-bit ALU, four full adders and four 4*1 multiplexers are used. Inverters are used in side B to invert the value for the subtraction operation. Full Adder is an adder that takes three inputs and outputs two results.

| Select $S_1$ | $S_0$ | $C_{in}$ | In $Y$ | Output $D = A + Y + C_{in}$ | Microoperation |
|---|---|---|---|---|---|
| 0 | 0 | 0 | $B$ | $D = A + B$ | Add |
| 0 | 0 | 1 | $B$ | $D = A + B + 1$ | Add w/carry |
| 0 | 1 | 0 | $\overline{B}$ | $D = A + \overline{B}$ | Subtract w/borrow |
| 0 | 1 | 1 | $\overline{B}$ | $D = A + \overline{B} + 1$ | Subtract |
| 1 | 0 | 0 | 0 | $D = A$ | Transfer A |
| 1 | 0 | 1 | 0 | $D = A + 1$ | Increment A |
| 1 | 1 | 0 | 1 | $D = A - 1$ | Decrement A |
| 1 | 1 | 1 | 1 | $D = A$ | Transfer A |

Fig. 11. 8 Operations of ALU

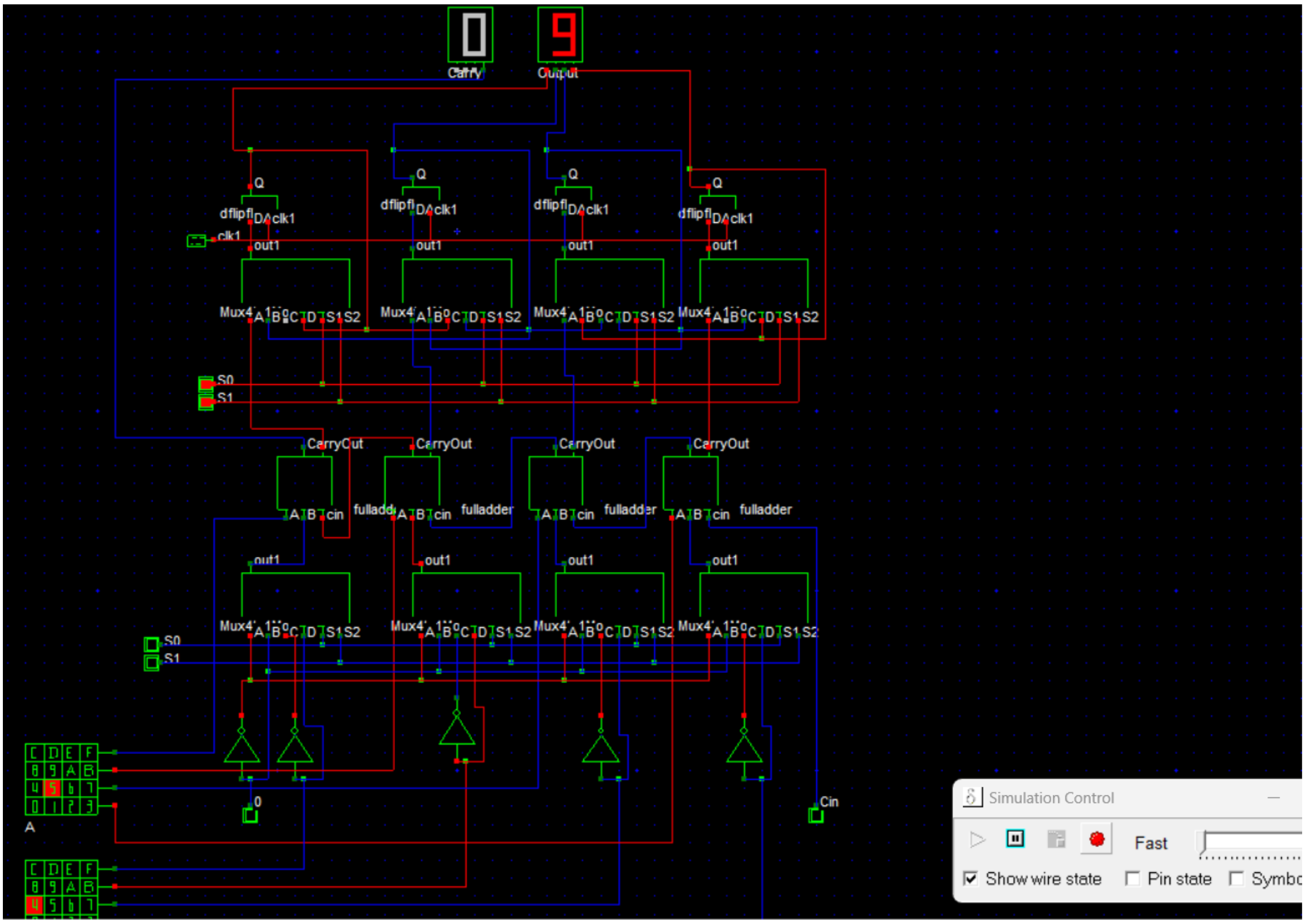

Fig. 12. 4-bit Universal Shift Register with ALU (addition)

In Fig. 12. the first two inputs are A and B, and the third is designated as Carry-in. The output carry is denoted as C-OUT, and the standard output is denoted as S or Sum. This ALU can perform eight microoperations, which the multiplexer's control pin can control. The microoperations are as follows: The whole system design is shown below.

## VI. Results and Discussion

Fig. 13 to 16 shows the time diagrams for all universal shift registers. The per division is for 1ns. Based on the input and output response, the delay can be defined.

Table 3: Delay comparison for different flip-flop

| Universal Shift Registers | Speed (Delay-output time-input time) |
|---|---|
| D flip flop | 9ns |
| Dynamic TGMS | 14ns |
| MTSPC double triggered | 22ns |
| TGMS | 9ns |

Table 3 and figure (from Fig 13 to 16) show that the D flip-flop and TGMS flip-flop-based shift register have less delay than the other flip-flops. However, the difference is that, unlike the dynamic TGMS flip-flop, it is much closer to the TGMS and D flip-flops. MTSPC has some other advantages but has a higher delay than others. After adding ALU to the D flip-flop-based universal shift register (Figs. 17 and 18), the delay increased slightly but not significantly. ALU was added to the TGMS flip-flop-based universal shift register, but the delay doubled in this perspective. So, overall, the D flip-flop is a better choice, but TGMS is also faster.

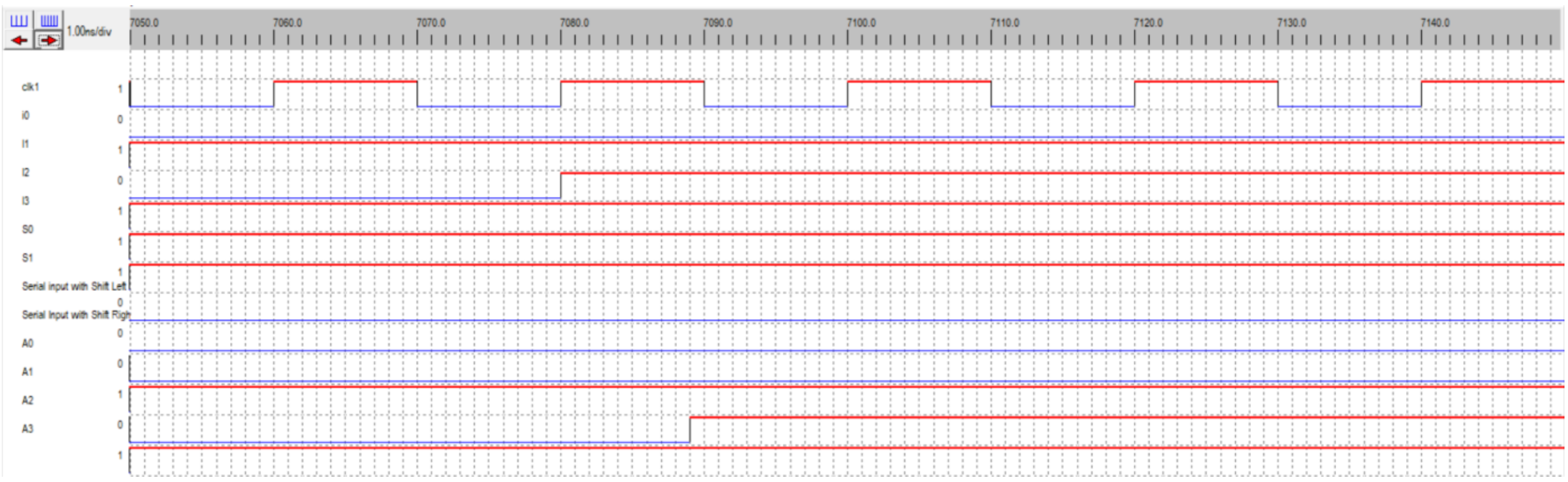

Fig. 13. Time Diagram of D flip flop-based universal shift register

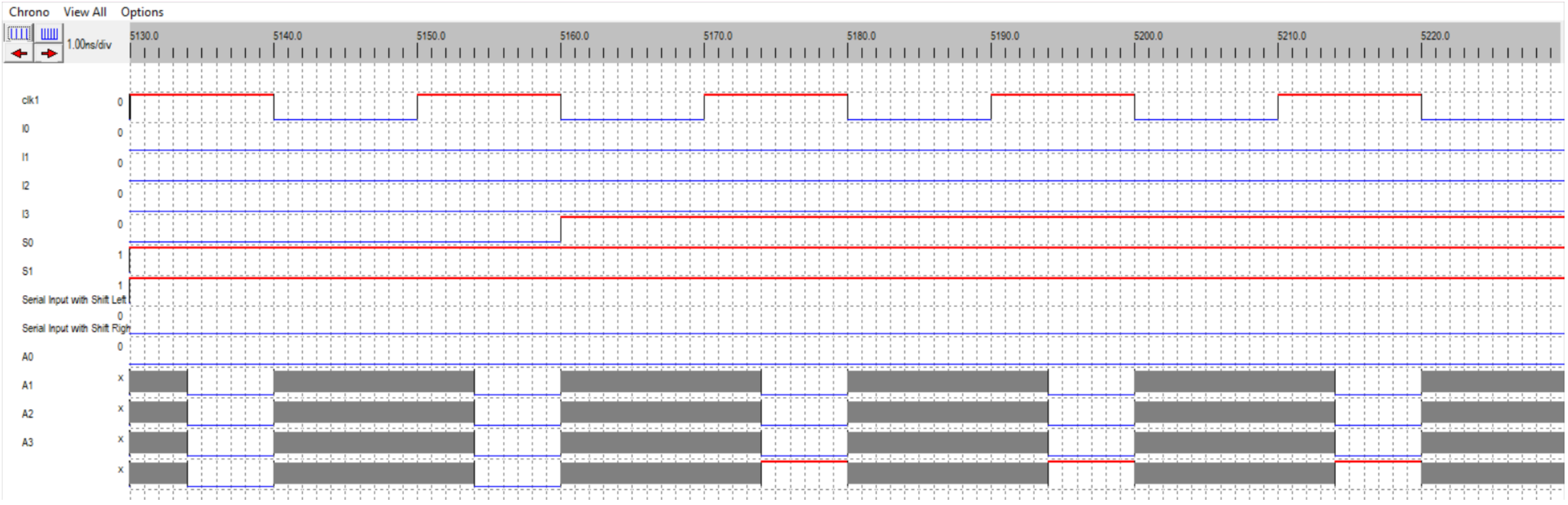

Fig. 14. Dynamic TGMS shift Register Time Diagram

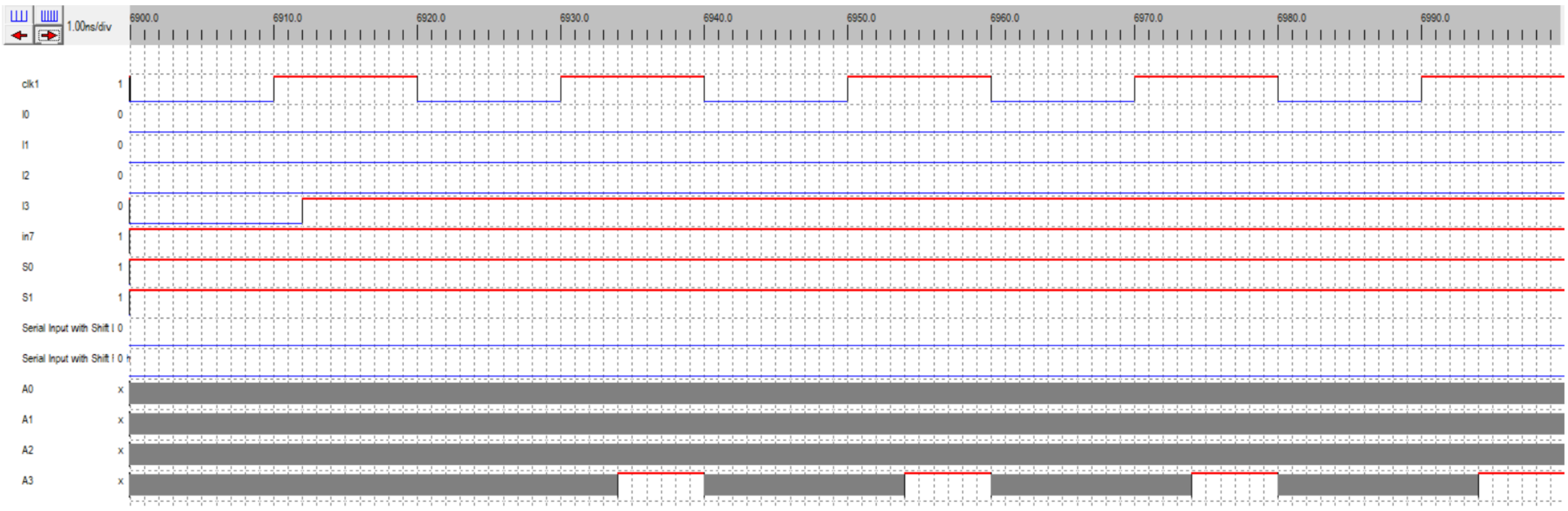

Fig. 15. MTSPC-based Universal Shift Register Time Diagram.

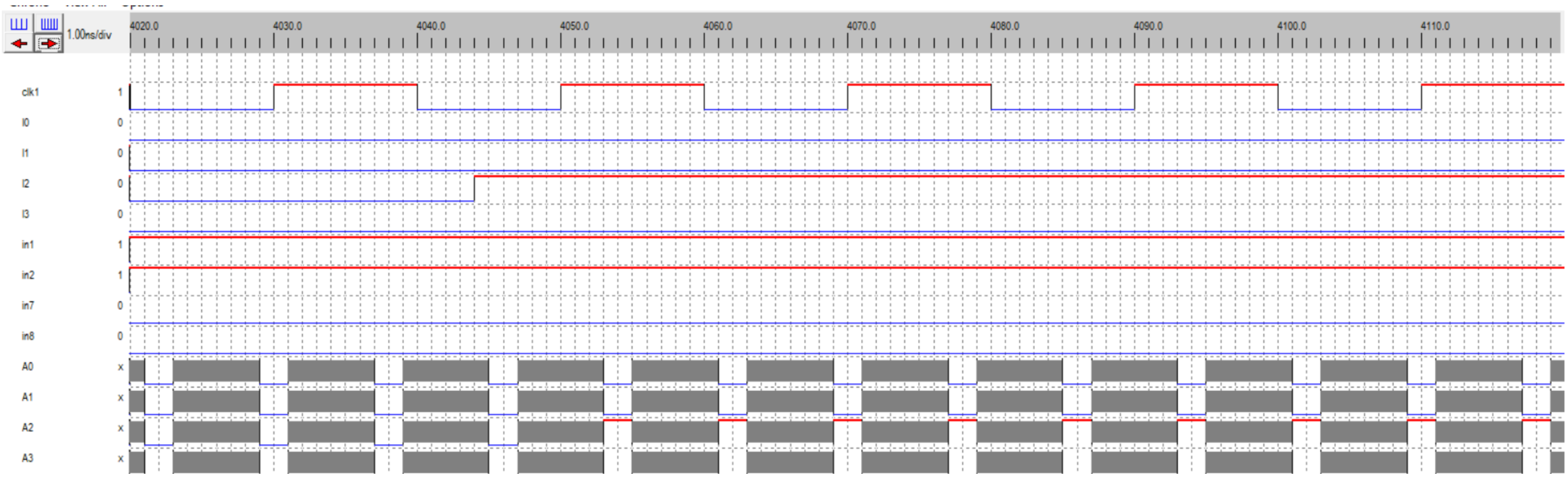

Fig. 16. TGMS D flip flop-based timing diagram

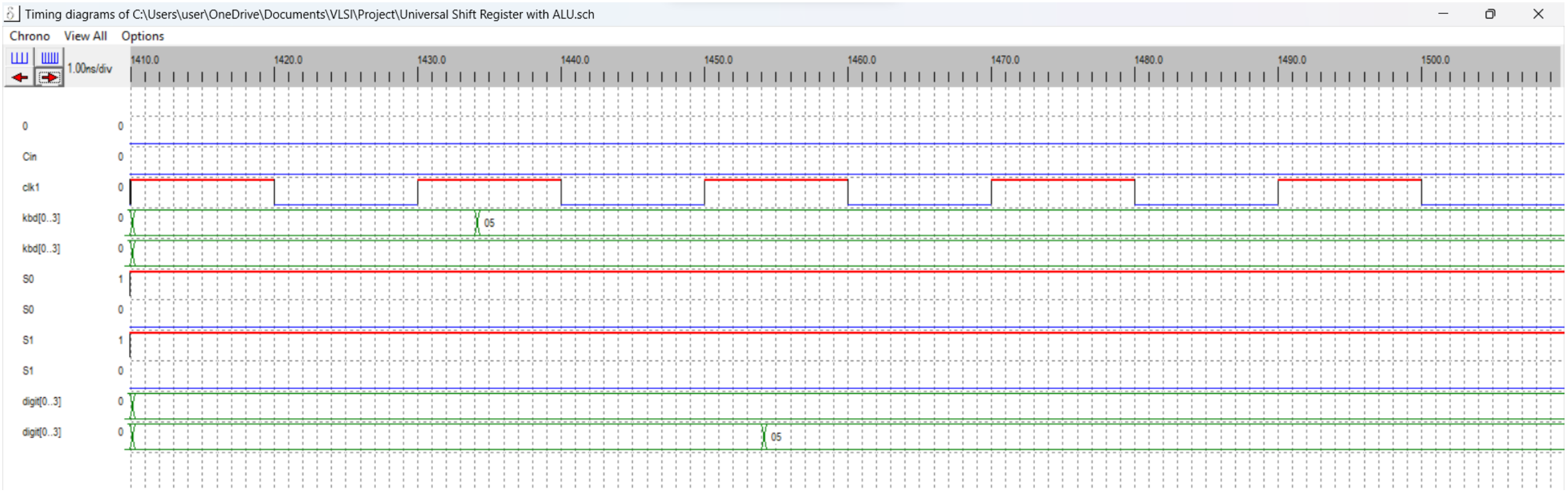

Fig. 17. Time Diagram for ALU (input)

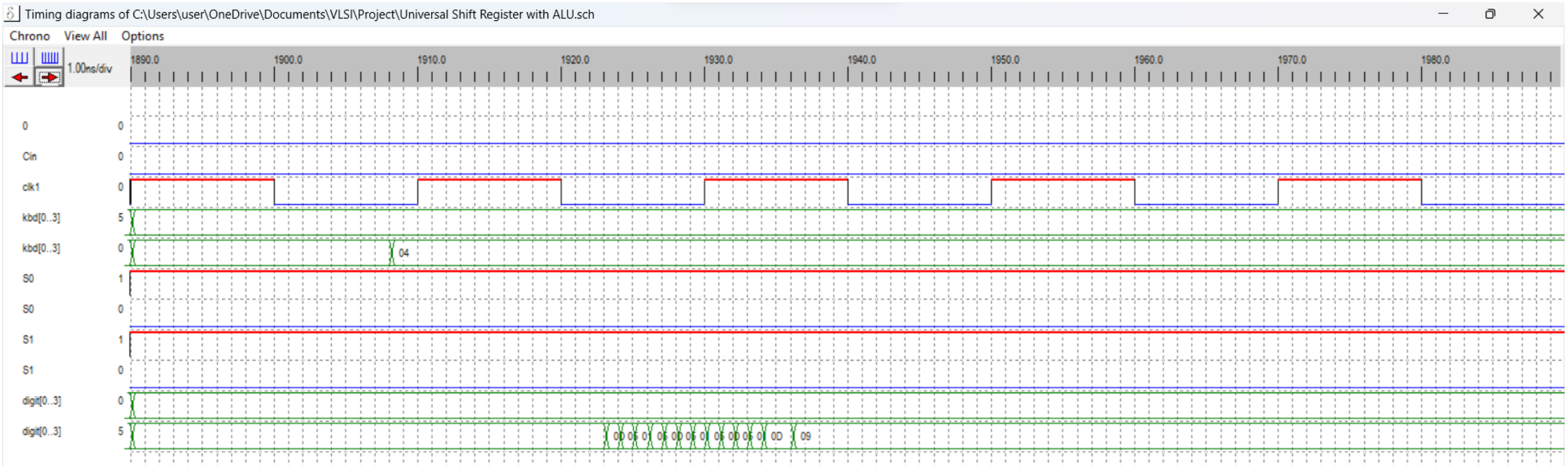

Fig. 18. Time diagram for ALU (for input & output)

## VII. FUTURE WORK

Though few types of flip-flops have been compared throughout the project, many scopes are still available to increase the speed of flip-flops. If there is a chance to implement artificial intelligence with this memory system to optimize design parameters, predict performance, or automate testing, it will become more facile, faster, and better. After that, we can invent a system that consumes less power but has the same speed. As a result, it will help the country's economy. In the future, we can implement fewer area-based flip flops like TGMS in the memory device so that speed becomes high and the area becomes less.

## VIII. CONCLUSION

This paper was about designing a memory system like a universal shift register added with 4-bit ALU. As a result, a memory device was created to store data efficiently. After that, different types of flip flops were compared to identify which universal shift registers were faster, and we got our desired result. In summary, the flip flop and TGMS have less delay than other flip flops, which means they are more rapid.